# Semantic Communication Networks Empowered Artificial Intelligence of Things

Yuntao Wang
School of Cyber Science and Engineering, Xi'an Jiaotong University, China

*Abstract*—Semantic communication aims to facilitate purposeful information exchange among diverse intelligent entities, including humans, machines, and organisms. It emphasizes precise semantic transmission over data fidelity, striving for meaningful expression while optimizing communication resources for efficient information transfer. Nevertheless, extant semantic communication systems face security, privacy, and trust challenges in integrating AI technologies for intelligent communication applications. This paper presents a comprehensive survey of security and privacy threats across various layers of semantic communication systems and discusses state-of-the-art countermeasures within both academic and industry contexts. Finally, we identify critical open issues in this burgeoning field warranting further investigation.

## I. Introduction

Semantic communication [1], [2] refers to purposeful information exchange between intelligent agents, which can include humans, intelligent machines, or other organisms. Its primary goal is the accurate transmission of semantic information between communication agents, rather than focusing on fidelity to the original data or signal [3]. In contrast to traditional communication methods that prioritize data representation and accuracy, semantic communication emphasizes extracting and conveying semantic-level information from the information source with the purpose of "preserving meaning" [4]. This approach is promised to advance the future Metaverse [5], [6], digital twins [7], and autonomous driving [8]. Current communication technologies aim to transmit more data with fewer errors using minimal communication resources, whereas semantic communication strives to convey the maximum amount of semantics with the least communication resource cost, rapidly delivering the least ambiguous semantic content.

However, as these systems continue to evolve, they are confronted with significant security, privacy, and trust challenges when integrating artificial intelligence (AI) technologies into intelligent communication applications. Specifically, the integration of AI technologies, especially large models [9], into semantic communication systems has brought about a paradigm shift in how information is exchanged and processed. While this shift offers numerous advantages, it also introduces new vulnerabilities and potential threats.

One of the fundamental issues facing semantic communication systems is security [10]. With AI-driven communication, there is an increased risk of unauthorized access, data breaches, and advanced persistent threats [11]. These threats can manifest in various forms, such as eavesdropping on conversations, intercepting and manipulating messages, and even masquerading as legitimate entities. Therefore, the security of these systems needs to be robust enough to withstand such threats. Privacy is another critical concern in semantic communication systems [6]. As these systems process large volumes of data, they often handle sensitive and personal information. Improper handling or exposure of this data can result in privacy breaches, identity theft, and other serious consequences. Safeguarding individual privacy while enabling effective communication is a challenging balancing act. Trust is also a cornerstone of semantic communication systems. Users should have confidence that the systems they rely on will function as intended, without compromising their data or the integrity of the communication process. The introduction of AI elements raises questions about the trustworthiness of these systems. Concerns such as AI bias, the transparency of AI decision-making, and accountability become prominent in this context.

In this paper, we present a comprehensive survey of security and privacy threats in semantic communication systems and discuss state-of-the-art countermeasures within both academic and industry contexts. Our objective is to gain a holistic understanding of the challenges in maintaining the confidentiality and privacy of exchanged information in the evolving landscape of communication technologies. Additionally, this study delves into the cutting-edge countermeasures currently being developed and implemented in academic and industry domains to mitigate these threats effectively.

In the following, we first review existing AI-driven security solutions for semantic communications and their challenges in Section II. In Section III, we discuss current defense mechanisms for semantic transmission security. Section IV discusses the security/privacy threats and the corresponding defense approaches in data and knowledge sharing within semantic communication networks. Finally, Section V concludes this work with conclusions and points out future research directions.

## II. AI-driven security for Semantic Communications

### A. Deep Learning for Secure Semantic Communications

Semantic communication is susceptible to adversarial examples due to the vulnerability of DNN-based codecs, resulting in semantic noise that can gradually alter the conveyed data's intended meaning. Current defenses against adversarial attacks include input data transformation, detection-based defenses, adversarial detectors, defensive distillation, and adversarial training [12]. It is proven to be effective in mitigating adversarial attacks and enhancing the robustness of DNNs in semantic communication by incorporating adversarial samples into the training dataset [12], [13], thereby enabling the semantic encoder and decoder to better adapt to such attacks.

Erdemir *et al.* [14] use a pre-trained StyleGAN model for semantic communication, achieving robust performance even with dynamic background knowledge. Kang *et al.* [13] introduce a semantic noise attack to produce adversarial data, employing a semantic distance minimization mechanism to mitigate their impact. Additionally, Du *et al.* [12] propose a training-free defense approach leveraging the visual invariance of semantically altered images to ensure accurate semantic feature extraction. By applying Gaussian image blurring, they reduce semantic similarity between the original and adversarial images, avoiding the need for retraining codecs in semantic communication.

In distributed semantic communication with massive connections, attackers may introduce various adversarial perturbations into wireless channels to confuse the transceiver. Nan *et al.* [3] design an innovative approach to create semantic-oriented perturbations for physical-layer adversarial attacks during semantic signal transmission. Additionally, they put forth an adversarial training technique to improve the resilience of semantic communication against diverse adversarial physical perturbations and other attacks, such as PGM [15].

### B. Large Models for Secure Semantic Communications

Large models, such as GPT-4, offer significant advantages for secure semantic communications by enabling multi-source data fusion and context-aware information exchange. These models can effectively prioritize critical messages, enhancing communication efficiency in resource-constrained environments. For example, Du *et al.* [10] develop a secure semantic communication system empowered by a large model with low overheads. They utilize multi-modal prompts to accurately reconstruct source content in changing environments, combining covert communication and diffusion models to securely transmit the multi-modal prompts and regenerate images in an energy-efficient manner. Typical security, privacy, trust, and ethical issues are comprehensively reviewed in [9].

## III. TRANSMISSION SECURITY OF SEMANTIC COMMUNICATION

### A. Physical-Layer Security for Key Generation in Semantic Communications

In traditional cryptographic algorithms, the complex processes of key generation and sharing pose significant challenges, hindering high transmission speeds in semantic communications. Physical-layer secure transmission technology integrates communication and security within semantic communications, offering a pivotal solution. Physical-layer key generation (PLKG) technology provides semantic communication with a quantum attack-resistant and key transmission-free method for real-time encryption [16]. This technology exploits unique transmission medium characteristics, such as wireless channel fading, to generate secret keys between communication parties such as unmanned aerial vehicles (UAVs) [17].

Studies [18], [19] utilize PLKG techniques to securely and efficiently generate keys in semantic communication. In [18], a reconfigurable intelligent surface (RIS) is used to devise a novel PLKG mechanism for a higher key generation rate by harnessing the randomness of transmitter-receiver semantic drifts. Nevertheless, this method mainly considers dynamic environments characterized by user mobility and changing conditions, making it less applicable in static environments (e.g., indoor semantic communication).

To address this issue, Aldaghri *et al.* [19] introduce a low-complexity technique named induced randomness for resource-limited semantic communication. In this approach, legitimate parties independently create local randomness and combine it with the unique properties of the wireless channel, enabling high-rate public randomness and secret key generation. The employed semantic security measures and information-theoretic analysis ensure resistance to eavesdropping attacks. Additionally, PLKG can be seamlessly integrated into hybrid cryptographic schemes, as demonstrated in [16].

### B. Robust Semantic Transmission in Emergency Networks

Semantic transmission enhances communication efficiency in emergency networks by prioritizing critical messages and reducing bandwidth usage, which is crucial in resource-constrained and dynamic disaster environments [20], [21]. By ensuring that essential information is accurately and swiftly conveyed, robust semantic transmission supports timely decision-making and efficient disaster response, ultimately improving the effectiveness of relief efforts and saving lives. For instance, [22] presents a robust and energy-efficient data transmission strategy using permissioned blockchain and reinforcement learning, offering insights for designing robust semantic transmission in emergency networks.

Additionally, Su *et al.* [23] develop a collaborative air-ground networking framework combining UAVs and ground vehicles for rapid disaster information and material delivery. By further empowering UAVs and ground vehicles with semantic communication capabilities, emergency networks can achieve more precise and context-aware data exchange, leading to enhanced coordination and faster response time. Future challenges include maintaining the accuracy of semantic interpretation and securing communication channels against attacks.

## IV. DATA AND KNOWLEDGE SECURITY IN SEMANTIC COMMUNICATIONS

### A. Blockchain for Trust-free Data and Knowledge Management

Blockchain technology offers potential solutions for trust-free data and knowledge management, particularly in distributed and collaborative environments [20]. Its decentralized nature ensures data integrity and immutability, eliminating the need for intermediaries and enhancing transparency. Smart contracts automate data validation and enforce predefined rules, reducing the potential for human error and fraud [24]. The authors in [4] introduce a semantic communication framework based on blockchain for AI-generated content (AIGC) services, which addresses security challenges from malicious semantic data attacks and ensures authenticity. They propose a semantic attack mechanism based on training to produce adversarial samples with similar descriptors but different meanings. Simulation results demonstrate the effectiveness of the defense scheme in differentiating authentic and adversarial semantic data, thereby safeguarding AIGC services.

In [25], a practical solution for remote data integrity is introduced, proposing the concept of blockchain-based private PDP. This scheme ensures secure data storage by leveraging blockchain technology. Key contributions include formalizing system and security models, designing a secure blockchain-based private PDP using RSA and quadratic residue groups, and providing efficiency analysis and implementation. The proposed scheme guarantees client anonymity, making it a robust solution for cloud storage integrity. Future research could explore identity-based and key-evolving blockchain-based PDP schemes to expand its applications. In [26], optimal energy-price contracts for different types of electric vehicles are deployed within smart contracts for trust-free distributed energy trading, demonstrating the versatility of blockchain technology in various applications. In [27], a blockchain-based semantic information exchange architecture is devised, where semantic information is tokenized into Non-Fungible Tokens (NFTs). This enables efficient and fair interactions through a Stackelberg game for pricing strategies, with zero-knowledge proof ensuring privacy-preserving trading, marking a substantial advancement over current NFT marketplaces.

However, the communication and computation overheads of the blockchain consensus process can be prohibitive for energy-constrained devices. To address these concerns, [28] designs a novel energy-recycling consensus protocol, along with redesigned blockchain structures and on-chain/off-chain collaboration mechanisms for improved performance. This protocol recycles wasted energy in proof-of-work (PoW)-series tasks by completing federated learning tasks, which are more useful for real-world services.

The social effects among device owners can be incorporated to develop efficient consensus protocols by leveraging trust relationships and collaborative behaviors to mitigate the impact of malicious actors and encourage active participation. By considering social factors such as reputation [17], mutual trust [29], and cooperation incentives [26], consensus mechanisms can be optimized to reduce latency, improve fault tolerance, and increase overall network efficiency.

Additionally, semantic computation tasks can be efficiently offloaded to edge nodes as they are closer to the data source, reducing latency and bandwidth usage. Edge nodes can process data locally, providing faster responses and easing the burden on central servers, promoting efficient, scalable, distributed processing. Existing offloading mechanisms based on reinforcement learning [30], stable matching game [31], and bargaining game [21] offer insights for semantic computation offloading.

B. Access Control of Semantic Data and Knowledge

Different access control (AC) policies, including role-based AC, attribute-based AC, content-based AC, purpose-aware AC, and mandatory AC, can safeguard data and knowledge in semantic communication based on specific requirements. For instance, in [32], the authors introduce a cryptographic scheme that combines Role-Based Encryption (RBE) with cloud storage systems. In the single-organization model (SO-RBE), encrypted data is accessible only to users meeting specific policies. The multi-organization model (MO-RBE) extends this capability, enabling cross-organization access based on users' roles. The design includes an efficient user revocation mechanism, reduces computational costs, and outsources cryptographic tasks to the cloud, minimizing user overhead. Security analysis proves resilience against Chosen Plaintext Attacks (CPA).

Xue et al. [33] introduce an advanced access control mechanism for outsourced data in the public cloud leveraging attribute-based encryption (ABE). This scheme enables collaborative access, allowing users to collaborate based on policies set by the data owner. It incorporates translation nodes in policy trees, ensuring secure collaboration and preventing unauthorized collusion attempts. The scheme embeds translation keys in secret keys and modifies ciphertexts, providing robust security features such as data confidentiality, controlled collaboration, and collusion resistance. Security analysis confirms its effectiveness, and performance evaluations demonstrate its practical viability for fine-grained access control in collaborative scenarios. The threat data sharing in AIoT [34] requires further investigation.

Xu et al. [35] present a privacy-preserving attribute-based access control strategy in cloud storage, utilizing a novel privacy-preserving and revocable ciphertext policy ABE mechanism. Additionally, it supports secure deduplication in cloud storage, enhancing efficiency. Security analysis demonstrates the CPA security, and the solution exhibits promising performance for integration into existing cloud storage services. To secure data sharing and analysis, Xue et al. [36] introduce a purpose-aware access control mechanism named SparkAC in Apache Spark. Two enforcement strategies, GuardSpark++ and GuardDAG, are proposed. GuardSpark++ optimizes Spark's logical query plans, ensuring security for relational queries, while GuardDAG handles all queries at the execution layer, introducing slightly higher overhead. Future work include the incorporation of obligations to enhance data owner awareness and control.

Additionally, blockchain enables fine-grained and automatic access control in semantic communication [37], [38]. For example, Saha et al. [38] design DHACS, a pioneering hybrid access control system based on the blockchain technology for industrial IoT. DHACS employs a smart contract framework based on three AC strategies: rule-based, role-based, and organization-based. Operational transactions from device controllers and their corresponding access controls are processed by a transaction pooler and block creator to generate validation-based access control smart contracts. DHACS addressesspecific operational dependencies within industrial IoT. Comprehensive experiments demonstrate DHACS's efficiency, showcasing a 30% reduction in complexity and 32.3% less energy consumption compared to existing solutions.

*C. Federated Learning for Privacy-Preserving Data and Knowledge Sharing*

Federated learning enables privacy-preserving data and knowledge sharing in collaborative semantic communications [6]. By allowing massive devices to collaboratively train a shared model locally without transferring private data, federated learning ensures sensitive information remains on-device, thereby enhancing privacy [39]. This decentralized approach mitigates the risks associated with centralized data storage and leverages distributed computational resources [40]. Federated learning facilitates the development of more accurate and context-aware semantic communication models while maintaining strict privacy.

Tong *et al.* [41] address audio semantic communication issues over wireless networks by proposing a model where edge devices transmit contextually significant audio data to a server. Utilizing a wav2vec-based autoencoder with convolutional neural networks (CNNs), the system ensures high-accuracy audio transmission with minimal data. Federated learning is applied across multiple devices and a server to enhance semantic information extraction, significantly reducing the mean squared error (MSE) of audio transmission and outperforming traditional coding schemes by nearly 100 times. For the immersive Metaverse constructed by billions of users with smart edge devices, Chen *et al.* [6] design a trustworthy semantic communication mechanism based on federated learning to securely handle multi-modal data while preserving privacy.

By leveraging trust and reputation among participating entities, federated learning models can be more efficiently trained, as entities are more likely to contribute high-quality data [42]. Social effects foster cooperative behavior, mitigating issues such as data poisoning and free-riding [43]. Additionally, incorporating social incentives can improve participation rates, leading to richer and more diverse datasets. This approach not only strengthens data privacy and security but also boosts the robustness and accuracy of semantic communication systems, ultimately enhancing collaborative efforts in distributed environments. Future research directions include robust federated learning paradigms [39], free-rider resistance [44], and privacy-preserving multimodal knowledge fusion.

## V. Conclusion

Semantic communication represents a crucial endeavor in the realm of information exchange, targeting the purposeful and accurate transmission of meaning among a broad spectrum of intelligent entities. The prioritization of semantic precision over data fidelity underscores the importance of effectively conveying intended meaning. However, integrating AI technologies into semantic communication systems presents substantial challenges, particularly in terms of security, privacy, and trust. This paper has thoroughly investigated these challenges, presenting a comprehensive survey of the security and privacy threats inherent in various layers of semantic communication systems. Both academic and industrial domains have made remarkable advancements in devising countermeasures to mitigate these threats, showcasing collective efforts to bolster the security and privacy of semantic communication systems.

Despite the progress made, several critical open issues persist within this burgeoning field. Ethical implications of AI integration, scalability concerns to accommodate growing demands, interoperability challenges, user education, and the ever-evolving landscape of security threats all require dedicated research and attention. Addressing these issues is paramount to ensuring the continued development and enhancement of secure, private, and trustworthy semantic communication systems. Further research and collaboration are indispensable in shaping the future landscape of semantic communication, emphasizing the imperative of continued exploration and innovation in this domain.